\documentclass[aps,prl,showpacs,preprint,floatfix]{revtex4-1}
\usepackage{graphicx,amsmath,amssymb,txfonts} 

\newcommand*{\bfig}[0]{\begin{figure}[p]}
\newcommand*{\efig}[0]{\end{figure}}
\newcommand*{\bfigwide}[0]{\begin{figure*}}
\newcommand*{\efigwide}[0]{\end{figure*}}

\newcommand{\C}[0]{$^\circ$C}
\newcommand{\E}[1]{E_{\text{#1}}}
\newcommand{\la}[1]{\lambda_{\text{#1}}}

\newlength{\wholefigwidth}
\newlength{\smallfigwidth}
\newlength{\halfsmallfigwidth}
\newcommand{\Fig}[1]{Fig.~\ref{fig:#1}}
\newcommand{\Tab}[1]{Table~\ref{tab:#1}}

\newcommand{\Eqn}[1]{Eqn.~\ref{eqn:#1}}

\newcommand*{\abinit}[0]{first-principles}
\newcommand*{\Abinit}[0]{First-principles}

\begin{document}
\title{Direct diffusion through interpenetrating networks: Oxygen in titanium}

\author{Henry H. Wu}
\author{Dallas R. Trinkle}
\email{dtrinkle@illinois.edu}
\affiliation{Department of Materials Science and Engineering, University of Illinois, Urbana-Champaign, Illinois 61801, USA}

\date{May 9, 2011}
\begin{abstract}
How impurity atoms move through a crystal is a fundamental and recurrent question in materials.  The previous understanding of oxygen diffusion in titanium relied on interstitial lattice sites that were recently found to be unstable, making the diffusion pathways for oxygen unknown.  Using first-principles quantum-mechanical methods, we find three oxygen interstitial sites in titanium, and quantify the multiple interpenetrating networks for oxygen diffusion.  Surprisingly, no single transition dominates, but all contribute to diffusion.
\end{abstract}

\maketitle

Controlling diffusion of impurity atoms into metals is a basic and ubiquitous technique of materials design for millennia.  The diffusion of oxygen in titanium impacts design of implant and aerospace alloys, as well as the formation of titanium-oxides.  Increasing the oxygen content in titanium forms ordered layered-oxide phases, which rely on the diffusion of oxygen into alternating basal planes to form\cite{Yamaguchi:1969xy, Tsuji:1997qv, Vykhodets:1997nr}; modeling the kinetics of ordering\cite{Ruban:2010jk} needs information about diffusion.  Initial stages of growth of titania nanotubes---e.g., for dye-sensitized solar cells---via anodization of a titanium metal substrate\cite{Macak:2005nt} involves the diffusion of oxygen.  At moderate to high temperatures, oxygen diffuses into titanium from the surface oxide, making the metal brittle, and requires costly processing steps to avoid\cite{Titanium2007}.  Designing titanium alloys with lower innate oxygen diffusivity has the potential to replace heavier alloys in aerospace to reduce greenhouse-gas emissions.  Ultimately, understanding how to impede or accelerate the diffusion of oxygen requires a fundamental description of diffusion pathways through titanium.

Diffusion of single oxygen atoms through hexagonal-closed packed ($\alpha$) titanium initially appears simple---proposed as atom-hopping between identical interstitial sites, following an Arrhenius relationship with temperature---but that simplicity hides a complex network of transition mechanisms.  Oxygen prefers to occupy an octahedral interstitial site surrounded by six titanium atoms\cite{Conrad:1981}, and so modeling oxygen diffusion had assumed either direct octahedral-to-octahedral transitions through tetrahedral transition states\cite{Vykhodets:1994fp}, or from octahedral to metastable tetrahedral sites\cite{Bertin:1980db}.  However, the recent discovery that the tetrahedral site is unstable in favor of a metastable hexahedral site\cite{Hennig:2005jk} left an open question: how does interstitial oxygen diffuse through $\alpha$-titanium?  Moreover, the ratio of oxygen diffusivity along basal ($xy$) directions and the $c$-axis ($z$) direction is nearly unity\cite{Vykhodets:1994fp} despite no symmetry relationship between the basal plane and the $c$-axis.

We explain these phenomena with \abinit\ predictions of \textit{three} distinct interstitial oxygen sites in $\alpha$-titanium---including a new non-basal crowdion site with lower symmetry than the octahedral and hexahedral sites---that create three different lattices interconnected through a network of transitions.  We derive the first entirely \abinit\ prediction of oxygen diffusion in $\alpha$-titanium, and our predictions compare well with experiment.  Our results explain the near isotropy of oxygen diffusion as transitions occur between the three different interstitial sites with similar absolute rates, so that nearly all portions of the transition network contribute to diffusion, rather than a single rate-controlling step.  That is, all escape times contribute to diffusion, not just from the octahedral site.  The metastable crowdion site and hexahedral site give a route for accelerating oxygen diffusion.  The full topology of the oxygen diffusion network highlights an unusual complexity for the kinetics of interstitials in metals.

The \abinit\ calculations are performed with \textsc{vasp}\cite{Kresse:1993jk,Kresse:1996xy}, a plane-wave density-functional theory (DFT) code.  Ti and O are treated with ultrasoft Vanderbilt type pseudopotentials\cite{Vanderbilt:1990qv,Kresse:1994nr} and the generalized gradient approximation of Perdew and Wang\cite{Perdew:1992ty}.  We use a single oxygen atom in a 96-atom ($4\times 4\times 3$) titanium supercell with a $2\times 2\times 2$ $k$-point mesh.  A plane-wave cutoff of 400eV is converged to 0.3meV/atom and the $k$-point mesh with Methfessel-Paxton smearing of 0.2eV is converged to 1meV/atom\cite{Hennig:2005jk}.  Projector augmented-wave (PAW) pseudopotential\cite{Kresse:1999jk} calculations with the PBE generalized gradient approximation\cite{Perdew:1996xy} give similar values, with a maximum error of 0.1eV (see supporting Table~1). From changes in supercell stresses for oxygen in different sites, we estimate the finite-size errors to be $\lesssim 0.05\text{eV}$; this is similar to the error found by using different computational cell sizes\cite{Hennig:2005jk}.  We use the climbing-image nudged elastic band\cite{Mills:1994yq,Henkelman:2000jk} method with one intermediate image and constant cell shape to find the transition pathways and energy barriers between different interstitial sites.  Along the path, the force is negated, while components perpendicular to the path are unchanged; the image relaxed to an extremum where the forces are less than 5meV/\AA, and restoring forces confirm that this extremum is a first-order saddle point.  The attempt frequency prefactor for each transition is estimated with the Vineyard equation\cite{Vineyard:1957vn}.  Only the restoring forces on the oxygen atom is used to compute the normal mode frequencies.  We use the three vibrational modes from the initial state and the two highest modes at the transition state to estimate the prefactor.  This does not include the coupling to and softening of neighboring Ti atoms; this underestimates the prefactors by less than 25\%%
\footnote{See EPAPS Document No. E-PRLTIO-XX-XXXXXXX for additional computational details for the basal and $c$-axis diffusion and the electronic density of states for oxygen at different sites.  This document may be found in the online article's HTML reference section, via the EPAPS homepage (http://www.aip.org/pubservs/epaps.html), or from ftp.aip.org in the directory /epaps/.  See the EPAPS homepage for more information.}.

\begin{figure}
\centering
\begin{tabular}{ccccc}
Site	&Wyckoff pos.	&$R_{\text{nn}}$ [\AA]	&$Z$	&$\Delta E$ [eV]\\
\hline
octahedral	&$2a$ $(0,0,0)$	&2.09	&6	&+0.00\\[3pt]
hexahedral	&$2d$ $(\frac23,\frac13,\frac14)$	&1.92	&5	&+1.19\\[3pt]
crowdion	&$6g$ $(\frac12,0,0)$	&2.00	&6	&+1.88
\end{tabular}

\includegraphics[width=0.75\smallfigwidth]{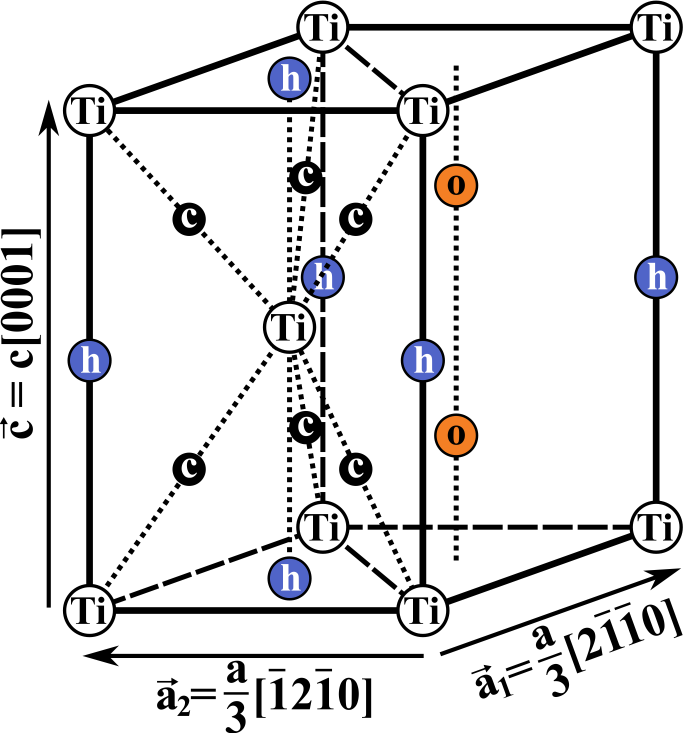}

\caption{Wyckoff positions\cite{ITC:A} and relative energies for oxygen interstitial sites in $\alpha$-titanium.  The interstitial site energy is reported relative to the octahedral site energy.  The geometry of each site is characterized by the nearest neighbor distance after relaxation, $R_{\text{nn}}$, with the coordination number $Z$.  Titanium atom sites are in white, while oxygen interstitial sites are in orange (octahedral), blue (hexahedral), and black (crowdion); the octahedral site is the ground state, with hexahedral and crowdion having site energies $\Delta E$ above.  The octahedral sites make a simple hexagonal lattice; the hexahedral sites, a hexagonal closed-packed lattice; and the crowdion sites, a kagom\'e lattice.}
\label{fig:sites}
\end{figure}

\Fig{sites} shows the hexagonal closed-packed unit cell of $\alpha$-titanium and the three interstitial sites for oxygen.  The crystal has space group 194, $P6_3/mmc$\cite{ITC:A}, where the crystal basis $\vec a_1$ and $\vec a_2$ are at an angle of 120$^\circ$ to each other in the hexagonal (``basal'') plane with length $a_{\text{Ti}}=2.933\text{\AA}$, while the $\vec c$ axis is perpendicular to both with length $c_{\text{Ti}}=4.638\text{\AA}$, and two titanium atoms per cell.  The octahedral (o) site is the equilibrium configuration for oxygen and is surrounded by 6 titanium atoms in a symmetric arrangement, 2.09\AA\ away.  The o-sites form a hexagonal lattice with a $c$-axis that is half of the titanium lattice.  Atomic forces on oxygen in the unstable tetrahedral site displace it towards the basal plane, into the hexahedral site\cite{Hennig:2005jk}; the hexahedral (h) site is 5-fold coordinated and is 1.19eV higher in energy than the o-site.  The three nearest titanium neighbors of the h-site are in the basal plane and are displaced to a distance of 1.92\AA, with two other neighbors directly above and below.  The h-sites form another hexagonal closed-packed lattice as $\alpha$-titanium with a translation of $[000\frac12]$.  The non-basal crowdion (c) site is 6-fold coordinated, but with lower symmetry and higher energy (1.88eV) than the o-site.  The two titanium atoms that contain each c-site have been significantly displaced such that all six of the c-site's titanium neighbors are approximately 2.00\AA\ away.  The c-sites form a kagom\'e lattice\cite{Syozi:1951jk} in the basal plane and is repeated along the $c$-axis twice per titanium unit cell.  The lowered symmetry of the c-sites in the kagom\'e lattice mean that a distortion of the unit cell can give the different c-sites \textit{different} energies.  We also considered a crowdion site in the basal plane, which is unstable.  A high formation energy is required to displace the two titanium atoms into the close-packed directions in the basal plane, while the two titanium neighbors of the non-basal crowdion can move in the softer pyramidal plane.  Note also that oxygen retains its divalency in all three configurations, and in the transition states (see supporting Fig.~1).

\begin{figure}
\centering
\includegraphics[width=\smallfigwidth]{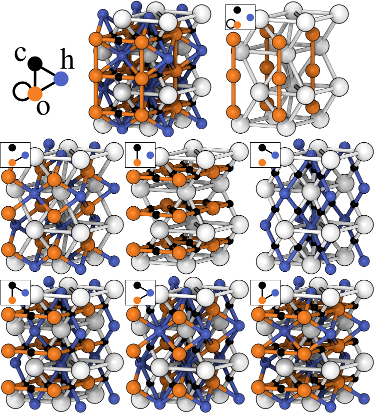}

\caption{Oxygen interstitial sites and oxygen diffusion pathways in $\alpha$-titanium.  White spheres are titanium atoms, orange spheres are octahedral interstitial sites, smaller blue spheres are hexahedral sites, and the smallest black spheres are crowdion sites.  The full transition pathway network is in the upper left, and is a superposition of all the remaining subnetworks.  The individual transition networks are o$\leftrightarrow$o, o$\leftrightarrow$h, o$\leftrightarrow$c, and h$\leftrightarrow$c.  The bottom three networks are formed from pairing up transition networks o$\leftrightarrow$h, o$\leftrightarrow$c, and h$\leftrightarrow$c.  All six networks contribute to the diffusion of oxygen.}
\label{fig:path}
\end{figure}

\Fig{path} show the interpenetrating network of transition pathways for oxygen between interstitial sites.  There are two out-of-plane transitions (with rate $\la{oo}$) from each o-site with its two direct neighbors in the $c$-axis.  The o-site is also surrounded by six h-sites and six c-sites and can transition into them (with rates $\la{oh}$ and $\la{oc}$).  The h-site is surrounded by six o-sites and six c-sites and can transition into them (with rates $\la{ho}$ and $\la{hc}$).  Each c-site resides in the center of a shared edge between two o-sites and two h-sites and can transition into them (with rates $\la{co}$ and $\la{ch}$).  With the exception of the o$\leftrightarrow$o $c$-axis transition, all other pathways are heterogeneous, starting and ending at different site types, and have not been considered previously.  The transition displacements and surrounding site symmetries are listing in \Tab{barriers}.

\begin{table}
\caption{Transition pathways, prefactors $\nu$, and energy barriers $E$ for oxygen diffusion in $\alpha$-titanium, between octahedral (o), hexahedral (h), and crowdion (c) sites.  The direction indicates the possible displacement vectors for the transition; the remaining transition vectors can be found by applying the point group symmetry operations.  The symmetries are listed in Hermann-Mauguin notation: $m$ is a mirror operation through the basal plane (0001), $\bar{1}$ is inversion, $\bar{3}$ is a 3-fold rotation axis around the $c$-axis [0001] with inversion, 6 is a 6-fold rotation axis around [0001], and $\frac3m$ is a 3-fold rotation axis around [0001] with mirror through (0001).  The absolute rate of transitions h$\to$x and c$\to$x is thermally activated with the transition energy barrier plus the site energy for h (+1.19eV) or c (+1.88eV), respectively; hence, all six heterogeneous transitions occur with similar absolute rates.}
\label{tab:barriers}

\centering
\begin{tabular}{@{}ccccc@{}}
&Direction	&Symmetry	&$\nu$ [THz]	&$E$ [eV]\\
\hline
o$\to$o	&$\langle000\frac12\rangle$	&$m$	&11.76	&3.25 \\[3pt]
o$\to$h	&$\langle\frac13\bar{\frac13}0\frac14\rangle$	&$\bar{3}$	&10.33	&2.04 \\[3pt]
o$\to$c	&$\langle\frac16\frac16\bar{\frac13}0\rangle$	&6	&16.84	&2.16 \\[3pt]
h$\to$o	&$\langle\frac13\bar{\frac13}0\frac14\rangle$	&$\frac3m$	&5.58	&0.85 \\[3pt]
h$\to$c	&$\langle\frac160\bar{\frac16}\frac14\rangle$	&$\frac3m$	&10.27	&0.94 \\[3pt]
c$\to$o	&$\langle\frac16\frac16\bar{\frac13}0\rangle$	&$\bar{1}$	&12.21	&0.28 \\[3pt]
c$\to$h	&$\langle\frac160\bar{\frac16}\frac14\rangle$	&$\bar{1}$	&13.81	&0.24 \\[3pt]
\end{tabular}
\end{table}

\Tab{barriers} summarizes the symmetries and energetics of all possible transitions for oxygen diffusion in $\alpha$-titanium.  The transition rate from site $i$ to $j$ at temperature $T$ is Arrhenius: $\lambda_{ij} = \nu_{ij} \exp( -E_{ij}/k_{\text{B}}T )$, where $E_{ij}$ is the energy barrier and $\nu_{ij}$ is the attempt prefactor for the transition.  The barrier of the direct $c$-axis transition between o-sites---$\E{oo}$---is too high to occur at relevant temperatures.  However, the lower barrier o$\leftrightarrow$h transition also passes through a triangular face of three titanium atoms like the o$\leftrightarrow$o $c$-axis transition.  This is similar to the instability of basal crowdion sites: the triangular face for the o$\leftrightarrow$o $c$-axis transition requires more energy to displace titanium atoms in the close-packed basal plane.  The triangular face for the o$\leftrightarrow$h transition is in the softer pyramidal plane allowing for easier titanium atom displacement.  Excluding the o$\leftrightarrow$o $c$-axis transition, all remaining transitions occur at approximately the same frequency---there is no single rate-controlling diffusion mechanism.  As the probability of a site $i$ being occupied is proportional to $\exp(-\Delta E_i/k_\text{B}T)$ for site energy $\Delta E_i$, the absolute rate of transitions is proportional to $\exp(-(\Delta E_i + E_{ij})/k_{\text{B}}T)$.  The transition barriers from h- and c-sites are lower than from o-sites, but the occupancy probability for h- and c-sites are lower.  Adding the site energy for h (+1.19eV) and c (+1.88eV) to the corresponding transition barriers reveal that all transitions occur with a temperature dependence of about $\sim$2.1eV; hence, \textit{all} of the interpenetrating transition networks contribute to the diffusion of oxygen.

We derive the exact rate equations for $c$-axis and basal diffusion using the multistate diffusion formalism\cite{Landman:1979jk,Landman:1979xy}.  The case of a single oxygen atom diffusing in a perfect $\alpha$-titanium lattice is represented by the periodic unit cell in \Fig{sites} with ten internal states.  The connected network leads to basal and $c$-axis diffusion rates that are similar.  The full diffusion equations (see supporting Eqn.~S11 and S12) are simplified when the rates of escape from the hexahedral and crowdion sites are much faster than from the octahedral site; then,
\begin{eqnarray}
\label{eqn:simp-dx}
D_{\text{basal}} &=& a_{\text{Ti}}^2 \left[\phantom{0}\la{oh} + \frac34\la{oc} + \frac14\frac{\la{oh}}{\la{ho}}\la{hc} + 0\la{oo}\right]\\
\label{eqn:simp-dz}
D_c &=& c_{\text{Ti}}^2 \left[\frac38\la{oh} + 0\la{oc} + \frac38\frac{\la{oh}}{\la{ho}}\la{hc} + \frac14\la{oo}\right]
\end{eqnarray}
This assumes that the crowdion sites are able to thermalize so that there is no correlated hops from the crowdion sites to neighboring sites.  The contribution of the individual rates to diffusion is similar for $\la{oh}$, $\la{oc}$, and $\la{hc}$ terms; at 300\C, the contributions are in ratios of 13.3:1.45:1 for basal diffusion, and 3.3:0:1 for $c$-axis; at 600\C, 7.1:1.8:1, and 1.8:0:1; at 900\C, 5.3:2.0:1, and 1.3:0:1; and at 1200\C, 4.4:2.1:1, and 1.1:0:1.  In all cases, the rate $\la{oo}$ is significantly smaller, contributing only $\lesssim 10^{-4}$ at 1200\C.  Over the temperature range of interest, all of the heterogeneous networks contribute to the diffusion of oxygen.

As temperature increases, we do not expect the crowdion site to achieve thermal equilibrium; hence, an increasing fraction of the o$\to$c jumps will become correlated basal o$\to$o jumps, and h$\to$c jumps will become correlated h$\to$h jumps.  This high temperature behavior can be approximated by removing the crowdions as metastable states from the network, and using $\la{oc}$ as the rate for direct basal o$\to$o transitions (and similarly $\la{hc}$ for h$\to$h transitions).  Then, the diffusion rates are bounded above by
\begin{eqnarray}
\label{eqn:highT-dx}
D^{\text{high T}}_{\text{basal}} &\lessapprox& a_{\text{Ti}}^2 \left[\phantom{0}\la{oh} + \frac32\la{oc} + \frac12\frac{\la{oh}}{\la{ho}}\la{hc} + 0\la{oo}\right]\\
\label{eqn:highT-dz}
D^{\text{high T}}_c &\lessapprox& c_{\text{Ti}}^2 \left[\frac38\la{oh} + 0\la{oc} + \frac34\frac{\la{oh}}{\la{ho}}\la{hc} + \frac14\la{oo}\right]
\end{eqnarray}
At 1200\C, the high temperature \Eqn{highT-dx} is 41\%\ larger than \Eqn{simp-dx} and \Eqn{highT-dz} is 48\%\ larger than \Eqn{simp-dz}; at 300\C, the differences are only 16\%\ and 23\%.  This suggests a small underestimation of diffusion rates at the highest temperatures.

\begin{figure}
\centering
\includegraphics[width=\smallfigwidth]{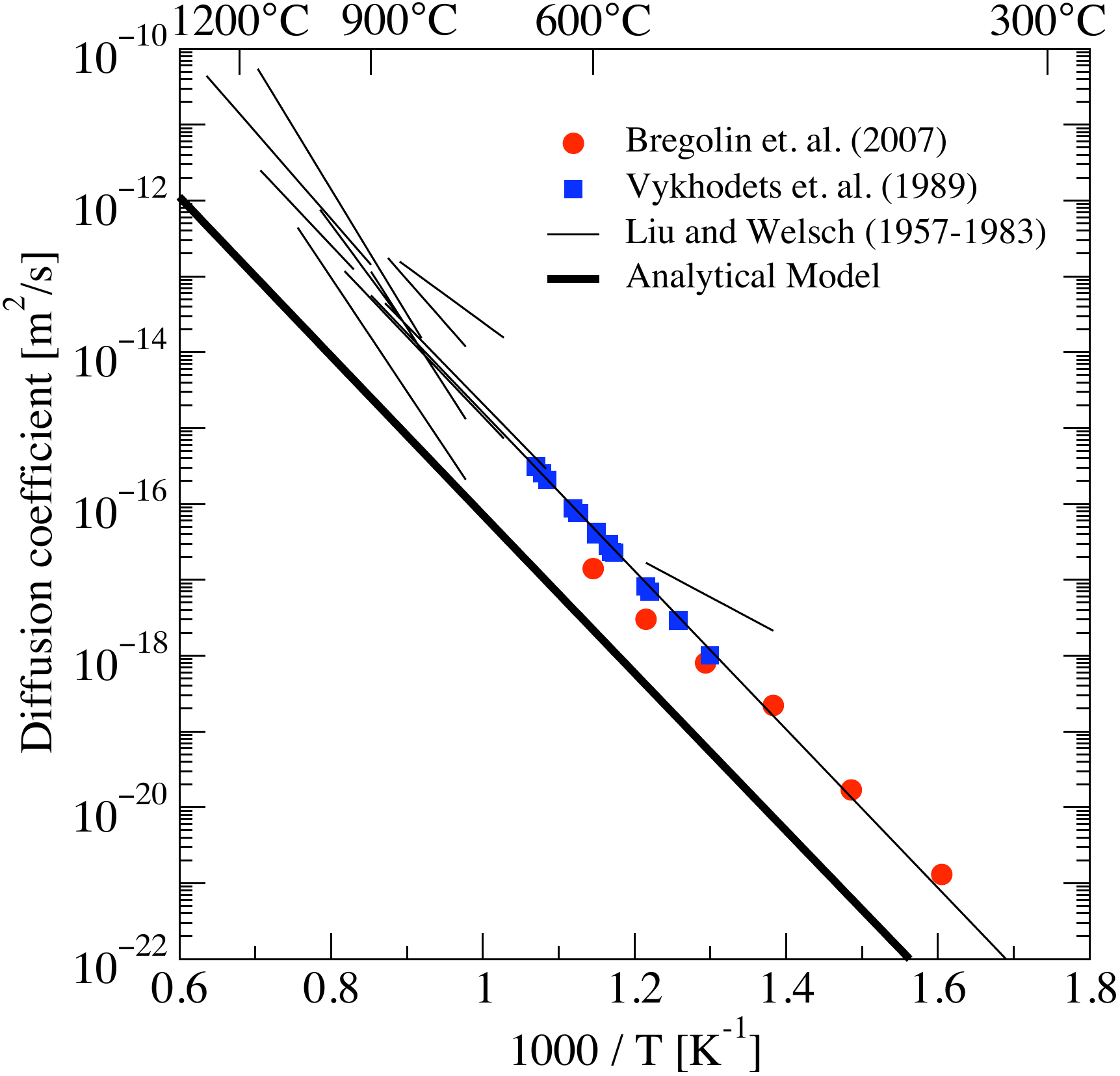}
    
\caption{Analytical results and experimental data of oxygen diffusivity in $\alpha$-titanium.  We compare our analytical DFT model (bold line) to experimental data from the literature survey by Liu and Welsch\cite{Liu:1988fv} (thin lines), and experiments by Bregolin\cite{Bregolin:2007eu} and Vykhodets\cite{Vykhodets:1989lq} (symbols).  Over the temperature range of 300--1200\C, the diffusion rate is Arrhenius $D_{0}$ = 2.18$\times$10$^{-6}$ $\text{m}^{2}\text{s}^{-1}$ with $E_{\text{act}}$ = 2.08eV.} 
\label{fig:diffusion}
\end{figure}

\Fig{diffusion} shows the diffusion coefficient from the multistate diffusion equations against experimental oxygen diffusion data.  From the diffusion equations \Eqn{simp-dx} and \Eqn{simp-dz}, the temperature behavior should follow the barriers, $\E{oh}$, $\E{oc}$, and $\E{hc}+\E{oh}-\E{ho}$.  An Arrhenius model with $D_{0}$ = $2.18\times10^{-6}\text{m}^{2}/\text{s}$ and a single barrier $E_{\text{act}}$ = 2.08eV matches \Eqn{simp-dx} and \Eqn{simp-dz} to within 15\%\ over the range 300--1200\C, with largest deviations at low temperatures.  The experimental data come from the literature survey\cite{Liu:1988fv} by Liu and Welsch and more recent experiments\cite{Bregolin:2007eu,Vykhodets:1989lq} using nuclear reaction analysis.  The activation energy matches well to experiment while the absolute diffusion coefficient is a factor of ten below experimental values---well within the expected accuracy of density-functional theory for diffusion.

\Abinit\ calculations determine the pathways for oxygen diffusion in $\alpha$-titanium, including a new interstitial non-basal crowdion site for oxygen in titanium.  Other than the high-barrier direct $c$-axis transition between octahedral sites, all transition paths are heterogeneous (o$\leftrightarrow$h, o$\leftrightarrow$c, and h$\leftrightarrow$c) and contribute to diffusion over a wide temperature range.  This shows that even well-studied materials science problems can have surprises: new configurations and new transitions give rise to complexity for single atom diffusion.  Moreover, the new sites suggest interesting interactions with titanium vacancies are possible, as they should destabilize nearby crowdion and perhaps hexahedral sites.  We expect other interstitial elements like carbon and nitrogen to have similar diffusion networks in titanium, and in other hexagonal-closed packed metals like magnesium and zirconium.  This new understanding of oxygen in titanium can serve as the basis for controlling oxygen diffusion in alloys, growth of oxide phases in titanium, and related challenges.

\begin{acknowledgments}
This research was supported by NSF/CMMI CAREER award 0846624 and Boeing.  The authors gratefully acknowledge the use of the Turing cluster maintained and operated by the Computational Science and Engineering Program at the University of Illinois.  The 3D models in \Fig{path} are visualized with VMD\cite{HUMP96} and rendered with Tachyon\cite{STON1998}.
\end{acknowledgments}

\end{document}